\documentclass[twocolumn,showpacs,preprintnumbers]{revtex4}
\usepackage{amsmath}
\usepackage{amssymb}
\usepackage{graphicx}
\usepackage{dcolumn}
\usepackage{bm}
\begin{document}

\title{Dark states of dressed spinor Bose-Einstein condensates of spin-1 atoms}
\author{Ma Luo, Chengguang Bao, Zhibing Li\footnote{%
Corresponding author: stslzb@mail.sysu.edu.cn}}
\affiliation{State Key Laboratory of Optoelectronic Materials and Technologies \\
School of Physics and Engineering \\
Sun Yat-Sen University, Guangzhou, 510275, P.R. China}

\begin{abstract}
Dressed states of spinor Bose-Einstein condensates of spin-1 atoms
located in optical cavity of near resonance frequency of excited
state of the atom are investigated. The wave function of dark states
are given.
\end{abstract}

 \maketitle

Implementation of optical trap of spin-1 to spin-3 bosonic atoms,
which librate the freedom of spin of the atoms, has given a great
chance to investigate more prolific physics of cold atomic cluster
\cite{01ho98,02kurn98,03stenger98,04gorlitz03,05griesmaier05}. When
temperature is low, spinor Bose-Einstein condensate (spinor BEC) is
form, whose eigenstate conserve the total spin and its Z-component
\cite{06bao04}. If the system is coupled to a high Q-factor optical
cavity mode, the eigenstate will be modified. This is a system of
cavity quantum electrodynamics (cavity QED) with a spinor BEC.

Cavity QED investigate the interaction between atom and optical
cavity mode \cite{07boca04,08manuz05}. In the parameter region that
the interaction strength between atom and optical mode is much
larger than decay rate of cavity mode and excited state of the atom,
the system is in strong coupling region. In this region, energy
transfer between atom and cavity mode by absorbing and emitting
photon periodically. The eigen state of the coupling system form
dressed state described by Jaynes-Cummings model \cite{09jaynes63}.
When there are more than one atom, the coupling effect will be
enhanced. Several previous research has investigated cavity QED with
a BEC theoretically and experimentally
\cite{10ferdinand07,11christoph05,12igor07,13peter00,14peter01,15elena98,16lee99}.
Dressed state of BEC is confirmed \cite{15elena98}. For a
multicomponent condensates, it is shown that dark states of dressed
BEC exist \cite{16lee99}. When the dressed BEC is in dark state, no
atom is state at excited state, but oscillate between two ground
states of $\Lambda$ type energy level atom. The research of dressed
BEC with has potential in understanding of quantum state
\cite{20higbie} and quantum information.

This paper consider the dark states of dressed spinor BEC of spin-1
atoms. At first, coupling Hamiltonian between spinor BEC and optical
cavity mode is given. And then, we calculate the matrix elements of
the coupling Hamiltonian, where we make use of fractional parentage
coefficients (FPCs) to get a simple analytic expression. At last,
wave function of the dark states is deduced and value scope of
quantum number for the dark states is fix.

\bigskip

The system we are going to investigate consist of $N$ $^{87}Rb$
atoms optically trapped to form a spinor BEC, located inside a
high-Q optical cavity. In the strong coupling region, the photon
decay rate of the cavity and the non-resonant decay rate of the
atoms is neglected. The optical trap can be modeled by a harmonic
trap potential. The frequency of the cavity mode is near resonance
to the transition frequency of $D_{2}$ line of $^{87}Rb$ :
$5^{2}S_{1/2},F=1\longrightarrow5^{2}P_{3/2},F=0$, so that only the
two levels is considered in the dressed states. The cavity consist
of two parallel high reflection mirror, which has two optical modes
at the resonance frequency: one mode is left circular polarization,
the other mode is right circular polarization. The two modes have
opposite angular momentum along the axis of the cavity. The
Hamiltonian of the atoms and cavity system can be written,
respective, as:
\begin{equation}
H_{A}=\sum_{i=1}^{N}(\frac{\mathbf{p}_{i}^{2}}{2m}+U(\mathbf{r}_{i})+\hbar\omega_{e}|e_{i}\rangle\langle
e_{i}|)+\sum_{i<j}V(\mathbf{r}_{i}-\mathbf{r}_{j}) \label{e1}
\end{equation}
\begin{equation}
H_{C}=\sum_{\mu=\pm 1}\hbar\omega_{c}\hat{a}_{\mu}^{+}\hat{a}_{\mu}
\end{equation}
where $\mathbf{p}_{i}$ is momentum of the ith atom, m is mass of the
atoms, $U(\mathbf{r})$ is harmonic potential of optical trap,
$|e_{i}\rangle$ is the excited state $5^{2}P_{3/2},F=0$ of the ith
atom, $\hbar\omega_{e}$ is energy level of the excited state,
$V(\mathbf{r}_{i}-\mathbf{r}_{j})$ is interaction between two atom
that depend on coupling of their spins, and $\hbar\omega_{c}$ is
energy level of cavity mode, $\hat{a}_{\mu}$ is annihilate operator
of standing wave cavity mode of polarization $\mu=\pm$.
$V(\mathbf{r}_{i}-\mathbf{r}_{j})$ can be given as
$V_{gg}(\mathbf{r}_{i}-\mathbf{r}_{j})|g_{i}g_{j}\rangle\langle
g_{i}g_{j}|+V_{ge}(\mathbf{r}_{i}-\mathbf{r}_{j})(|g_{i}e_{j}\rangle\langle
g_{i}e_{j}|+|e_{i}g_{j}\rangle\langle e_{i}g_{j}|)$, where
$|g_{i}\rangle$ is ground state of the ith atom. $V_{gg}$ is spin
dependent interaction between two ground state atoms with spin-1,
which is
$(c_{0}+c_{2}\mathbf{F}_{i}\cdot\mathbf{F}_{j})\delta(\mathbf{r}_{i}-\mathbf{r}_{j})$,
$\mathbf{F}_{i}$ is spin operator of the ith atom, $c_{0}$($c_{2}$)
is spin independent (dependent) interaction constant \cite{01ho98}.
For $V_{ge}$, one of the atom is spin-0, the scattering has only one
channel, so that the interaction is spin independent, i.e.
$V_{ge}=c_{4}\delta(\mathbf{r}_{i}-\mathbf{r}_{j})$ where $c_{4}$ is
a constant.

The direction of quantize axis of the atoms is along the axis of the
cavity, so that by absorbing a left (right) circular polarization
photon from the cavity, only the $\sigma^{+}$ ($\sigma^{-}$)
transition that conserve the angular momentum could occur; the same
conservation law of angular momentum apply for emission one photon
to the cavity. Therefore, the Hamiltonian of coupling between atoms
and cavity mode is,
\begin{equation}
H_{AC}=\sum_{i=1}^{N}[\hbar\Omega_{+}|e_{i}\rangle\langle
g_{i},-1|\hat{a}_{+}+\hbar\Omega_{-}|e_{i}\rangle\langle
g_{i},+1|\hat{a}_{-}]+H.c.
\end{equation}
where
$\hbar\Omega_{\pm}=d_{\pm}\sqrt{\hbar\omega_{c}/2\epsilon_{0}V}$ is
coupling strength of $\sigma^{\pm}$ transition, $d_{\pm}$ is the
corresponding atomic dipole-matrix element that satisfy
$d_{+}=-d_{-}$, $V$ is mode volume of the cavity mode, $|g_{i},\pm
1\rangle$ is the ground state of atom with spin component $\pm1$.

\bigskip

The eigenstate of non-coupling Hamiltonian $H_{A}+H_{C}$ can be
labeled as $|g,S,M;N_{1},N_{2}\rangle$ and
$|e,S',M';N_{1},N_{2}\rangle$, where $g$ means all N atoms are at
ground state, $e$ means one of the atoms is at excited state, $S$
and $M$ is total spin and the $Z$ component, $N_{1}$ ($N_{2}$) is
photon number of $\mu=+$ ($\mu=-$) polarization photon. For the
state that all N atoms are at ground state, $N-S$ is limited to be
even, and the eigen energy is approximated to be
$E_{g,S}^{N_{1},N_{2}}=J_{g}S(S+1)+\hbar\omega_{c}(N_{1}+N_{2})$
\cite{06bao04}, where $J_{g}$ is spinor interaction constant that
depend on total spin and spatial wave function. When one atom is at
excited state of spin-0, all the other atoms couple to total spin
$S'$, with the limitation that $N-1-S'$ is even. The atomic part of
wave function of this state can be written as
\begin{equation}
|e,S',M'\rangle=\frac{1}{\sqrt{N}}\sum_{i}\{[\varphi_{e}(\mathbf{r}_{i})\prod_{j\ne
i}\varphi_{g}(\mathbf{r}_{j})]\vartheta^{N-1}_{S',M'}(\bar{i})\}
\label{e4}
\end{equation}
where $\varphi_{e(g)}$ is spatial wave function of excited (ground)
state atom, $\vartheta^{N-1}_{S',M'}(\bar{i})$ is total symmetric
spin state of all the other atoms without the ith atom with quantum
number $S'$ and $M'$. Spin of the ith atom is zero, and it couple
with spin state of the other $N-1$ atoms, forming total spin state
$S'$. Using variation method to deduce Schrodinger equation for the
spatial wave function $\varphi_{e(g)}$ and solving the equation by
perturbation approximation, eigen energy of $|e,S',M'\rangle$ plus
energy of the cavity mode is given as
$E_{e,S'}^{N_{1},N_{2}}=J_{e}S'(S'+1)+\hbar\omega_{e}+\hbar\omega_{c}(N_{1}+N_{2})$.
$J_{e}$ is spinor interaction constant. When $N$ is small, $J_{g}$
and $J_{e}$ is approximated to be equated to each other, noted as
$J$. $J$ is much smaller than $\hbar\omega_{c}$ and
$\hbar\omega_{e}$, so that for a fix set of quantum number
$(g(e),N_{1},N_{2})$, the energy levels expand to be a narrow band
by quantum number $S(S')$ around energy level
$(\hbar\omega_{e})+\hbar\omega_{c}(N_{1}+N_{2})$. When there is no
static magnetic field, the energy levels do not depend on $M(M')$.
We consider the dressed states consist of states in the first
excited band: $|g,S,M;1,0\rangle$, $|g,S,M;0,1\rangle$ and
$|e,S',M';0,0\rangle$. According to the conservation of angular
momentum at quantize axis, the dressed states with total angular
momentum along quantize axis $M\hbar$ can be written as
\begin{eqnarray}
|M\rangle &=& \sum_{S'}c_{e}^{S',M}(t)|e,S',M;0,0\rangle \notag \\
& &+\sum_{S}c_{+}^{S,M-1}(t)|g,S,M-1;1,0\rangle \notag \\ &
&+\sum_{S}c_{-}^{S,M+1}(t)|g,S,M+1;0,1\rangle
\end{eqnarray}
where the summation of $c_{e}^{S',M}$ run through $S'$ that $N-S'-1$
is even and $S'>=M$, the summation of
$c_{+}^{S,M-1}$($c_{-}^{S,M+1}$) run through $S$ that $N-S$ is even
and $S>=M-1$($S>=M+1$).

\bigskip

In order to get the equation of motion of probability amplitudes
under the full Hamiltonian $H=H_{A}+H_{C}+H_{AC}$, we deduce the
matrix element of the coupling part of Hamiltonian $H_{AC}$ at the
bases of eigen states of non-coupling Hamiltonian. We consider
matrix element $\langle e,S',M';0,0|H_{AC}|g,S,M;1,0\rangle$ first.
The atomic state $\langle e,S,M|$ is expanded as equation
(\ref{e4}). The atomic spin state of $|g,S,M;1,0\rangle$ is total
symmetric spin state of $N$ atoms, $\vartheta^{N}_{S,M}$. Making use
of FPCs, the spin state of the ith atom can be extracted, so that
the atomic state can be expanded as
\begin{eqnarray}
|g,S,M\rangle=[\prod_{j=1}^{N}\varphi_{g}(\mathbf{r}_{j})] &
&[a_{S}^{N}[\chi(i)\vartheta^{N-1}_{S+1,M}(\bar{i})]_{S} \notag
\\+& &b_{S}^{N}[\chi(i)\vartheta^{N-1}_{S-1,M}(\bar{i})]_{S}]
\end{eqnarray}
where $a_{S}^{N}=\{[1+(-1)^{N-S}](N-S)(S+1)/[2N(2S+1)]\}^{1/2}$ and
$b_{S}^{N}=\{[1+(-1)^{N-S}]S(N+S+1)/[2N(2S+1)]\}^{1/2}$ is FPCs
given in reference \cite{17bao05}, $\chi(i)$ is spin state of the
ith spin-1 ground state atom. The coupling between $\chi(i)$ and
$\vartheta^{N-1}_{S\pm1,M}(\bar{i})$ can be expanded by using
Clebsch-Gordan coefficients \cite{18varshalovich88}. When we
calculate the matrix element of operator of the ith atom in
$H_{AC}$, we can extracted spin state of the ith atom and make use
of the orthogonal relation of spin states to get the exact
expression of the spin part of the element. Under approximation that
the spatial wave functions are independent of total spin and
internal state, the matrix element can be expressed as
\begin{eqnarray}
& &\langle e,S',M';0,0|H_{AC}|g,S,M;1,0\rangle \notag \\
&=&\sqrt{N}\hbar \Omega_{+}
a^{N}_{S}C^{SM}_{S',1,M',-1}\delta_{S,S'-1}\delta_{M,M'-1} \notag \\
&+&\sqrt{N}\hbar \Omega_{+}
b^{N}_{S}C^{SM}_{S',1,M',-1}\delta_{S,S'+1}\delta_{M,M'-1}
\end{eqnarray}
where $C^{SM}_{S_{1},S_{2},M_{1},M_{2}}$ is Clebsch-Gordan
coefficients. Using similar analysis, the matrix element with
another photon state can be obtained as
\begin{eqnarray}
& &\langle e,S',M';0,0|H_{AC}|g,S,M;0,1\rangle \notag \\
&=&\sqrt{N}\hbar \Omega_{-}
a^{N}_{S}C^{SM}_{S',1,M',1}\delta_{S,S'-1}\delta_{M,M'+1} \notag \\
&+&\sqrt{N}\hbar \Omega_{-}
b^{N}_{S}C^{SM}_{S',1,M',1}\delta_{S,S'+1}\delta_{M,M'+1}
\end{eqnarray}

\bigskip

Inserting the state into Schrodinger equation, and take the inner
product with all bases in equation (4), the equation of motion of
probability amplitudes are obtained as
\begin{eqnarray}
i\hbar\frac{dc_{e}^{S',M}}{dt}&=&[\hbar\omega_{e}+J
S'(S'+1)]c_{e}^{S',M}
\notag \\
&+&\sqrt{N}\hbar \Omega_{+} a^{N}_{S'-1}C^{S'-1,M-1}_{S',1,M,-1}c_{+}^{S'-1,M-1} \notag \\
&+&\sqrt{N}\hbar \Omega_{+} b^{N}_{S'+1}C^{S'+1,M-1}_{S',1,M,-1}c_{+}^{S'+1,M-1} \notag \\
&+&\sqrt{N}\hbar \Omega_{-} a^{N}_{S'-1}C^{S'-1,M+1}_{S',1,M,1}c_{-}^{S'-1,M+1} \notag \\
&+&\sqrt{N}\hbar \Omega_{-}
b^{N}_{S'+1}C^{S'+1,M+1}_{S',1,M,1}c_{-}^{S'+1,M+1} \label{e9}
\end{eqnarray}
where $N-S'-1$ is limited to be even, and
\begin{eqnarray}
i\hbar\frac{dc_{+}^{S,M-1}}{dt}&=&[\hbar\omega_{c}+J
S(S+1)]c_{+}^{S,M-1}
\notag \\
&+&\sqrt{N}\hbar \Omega_{+} a^{N}_{S}C^{S,M-1}_{S+1,1,M,-1}c_{e}^{S+1,M} \notag \\
&+&\sqrt{N}\hbar \Omega_{+}
b^{N}_{S}C^{S,M-1}_{S-1,1,M,-1}c_{e}^{S-1,M} \label{e10}
\end{eqnarray}
\begin{eqnarray}
i\hbar\frac{dc_{-}^{S,M+1}}{dt}&=&[\hbar\omega_{c}+J
S(S+1)]c_{-}^{S,M+1}
\notag \\
&+&\sqrt{N}\hbar \Omega_{-} a^{N}_{S}C^{S,M+1}_{S+1,1,M,1}c_{e}^{S+1,M} \notag \\
&+&\sqrt{N}\hbar \Omega_{-}
b^{N}_{S}C^{S,M+1}_{S-1,1,M,1}c_{e}^{S-1,M} \label{e11}
\end{eqnarray}
where $N-S$ is limited to be even. $M\hbar$ is total angular
momentum along quantize axis. The exist of dark state require that
the probability amplitude $c_{e}^{S',M}$ equate to zero. From
equation (\ref{e10}) and (\ref{e11}), this will result in the total
spin dependence of the time factor of $c_{\pm}^{S,M\mp1}$. From
equation (\ref{e9}), this will result in relations between different
probability amplitude. Because the time factor of
$c_{\pm}^{S,M\mp1}$ depend on $S$ as $e^{-iJS(S+1)t/\hbar}$, and
$\Omega_{+}=-\Omega_{-}$, equation (\ref{e9}) reduce to two equation
for dark states,
\begin{equation}
a^{N}_{S'-1}C^{S'-1,M-1}_{S',1,M,-1}c_{+}^{S'-1,M-1}-a^{N}_{S'-1}C^{S'-1,M+1}_{S',1,M,1}c_{-}^{S'-1,M+1}=0
\end{equation}
\begin{equation}
b^{N}_{S'+1}C^{S'+1,M-1}_{S',1,M,-1}c_{+}^{S'+1,M-1}-b^{N}_{S'+1}C^{S'+1,M+1}_{S',1,M,1}c_{-}^{S'+1,M+1}=0
\end{equation}

\bigskip

Thus far, it is easy to find that a dark state consist of two states
with the same total angular momentum along quantize axis, $M\hbar$,
and the same total spin $S$. The two states are
$|g,S,M-1;1,0\rangle$ and $|g,S,M+1;0,1\rangle$, and their
probability amplitude satisfy the equations
\begin{equation}
a^{N}_{S}C^{S,M-1}_{S+1,1,M,-1}c_{+}^{S,M-1}-a^{N}_{S}C^{S,M+1}_{S+1,1,M,1}c_{-}^{S,M+1}=0 \notag \\
\end{equation}
\begin{equation}
b^{N}_{S}C^{S,M-1}_{S-1,1,M,-1}c_{+}^{S,M-1}-b^{N}_{S}C^{S,M+1}_{S-1,1,M,1}c_{-}^{S,M+1}=0
\label{e14}
\end{equation}
Existence of solution for equation (\ref{e14}) is the determine of
coefficient matrix equation to zero, which gives
\begin{eqnarray}
&
&a_{S}^{N}b_{S}^{N}(C^{S,M-1}_{S+1,1,M,-1}C^{S,M+1}_{S-1,1,M,1}-C^{S,M+1}_{S+1,1,M,1}C^{S,M-1}_{S-1,1,M,-1})
\notag \\
&=&a_{S}^{N}b_{S}^{N}\frac{2M(2S+1)}{\sqrt{4S(2S-1)(S+1)(2S+3)}}=0
\end{eqnarray}
It is obvious that one kind solution is $M=0$ and $S>0$. When $S=0$,
$M$ could not be $\pm1$; when $S>0$ and $M=0$, all pairs of
($|g,S,-1;1,0\rangle$ , $|g,S,1;0,1\rangle$) form dark states with
$c_{+}^{S,-1}=c_{-}^{S,1}$. Therefore, this kind of dark state is
\begin{equation}
|S,M=0\rangle_{d}=\frac{1}{\sqrt{2}}(|g,S,-1;1,0\rangle+|g,S,1;0,1\rangle)\label{e16}
\end{equation}
Another kind of solution is $a^{N}_{S}=0$ when $S=N$. In this case,
there is not atomic excited state with total spin $S=N+1$, so that
only the second equation in (\ref{e14}) is required. Thus, when
$S=N$, all pairs of ($|g,N,M-1;1,0\rangle$ , $|g,N,M+1;0,1\rangle$)
form dark states with
$\sqrt{(N-M)(N-M+1)}c_{+}^{S=N,M-1}=\sqrt{(N+M)(N+M+1)}c_{-}^{S=N,M+1}$.
This kind of dark state is
\begin{eqnarray}
&
&|S=N,M\rangle_{d}=\frac{\sqrt{(N+M)(N+M+1)}}{\sqrt{2(M^{2}+N+N^{2})}}|g,N,M-1;1,0\rangle\notag
\\
&
&+\frac{\sqrt{(N-M)(N-M+1)}}{\sqrt{2(M^{2}+N+N^{2})}}|g,N,M+1;0,1\rangle
\label{e17}
\end{eqnarray}
In equation (\ref{e16}) and (\ref{e17}), the time factor in include
into the state itself, because it only depend on $S$.

In the above analysis, it is assumed that magnetic field is zero.
However, it is impossible to have absolute zero magnetic field
environment. Therefore, we consider how small the magnetic field is
required to achieve dark states. Because the dark states consist of
two states with atomic Z-component of angular momentum difference of
$2\hbar$, the Zeeman split is $2gB$, where $B$ is magnetic field,
$g=-0.70MHz/G$ is Zeeman split factor \cite{19hans97}. The energy
difference between state with $S+1$ and $S-1$ total spin is
$J(4S+2)$. For a system with $N=1000$ atoms and
$\omega_{trap}=1000Hz$ optical trap frequency, $J$ is estimated to
be $1Hz$ approximately. if Zeeman split is much smaller than energy
split due to total spin, .i.e. $2gB<<J(4S+2)$, the Zeeman split can
be neglected and the dark state exist; otherwise, the dark states is
damaged. It can be seen that magnetic field is require to be smaller
than $10^{-6}G$ so as to mean the requirement of dark states. Since
magnetic field on the surface of the earth is about $0.6G$, it need
very accurate experiment set up the counteract this magnetic field.

\bigskip

In conclusion, the Hamiltonian of dressed spinor BEC regardless of
atomic and photon decay is given. The Hamiltonian is diagonalized
under bases of eigenstates of non-coupling part of Hamiltonian of
spinor BEC and optical cavity mode. Only the matrix elements that
conserve total angular momentum along quantize axis is nonzero. As a
result, the dressed state with total angular momentum along quantize
axis $M\hbar$ consist of three kind of states in non-coupling
bases:$|e,S',M;0,0\rangle$, $|g,S,M-1;1,0\rangle$ and
$|g,S,M+1;0,1\rangle$. Dark state is a state that oscillate between
two kind of states, $|g,S,M-1;1,0\rangle$ and $|g,S,M+1;0,1\rangle$,
while is not excited to the states $|e,S,M;0,0\rangle$. It is found
that there are two species of dark states, whose wave function is
given by equation (\ref{e16}) and equation (\ref{e17}),
respectively. It is obvious from the two wave functions that the
dark state is entangle state of spinor BEC $|g,S,M\pm1\rangle$ and
optical cavity mode $|\mu=\pm\rangle$. This entanglement could be
very useful to the theory and application of quantum information and
quantum computer.

\bigskip

\begin{acknowledgments}
We appreciate the support from the NSFC under the grants 10574163 and
10674182.
\end{acknowledgments}

\clearpage

\end{document}